\documentclass[aip,rsi,reprint]{revtex4-1}
\usepackage{graphicx}
\usepackage{amsmath,amsthm,amssymb,latexsym,amsfonts,times,mathrsfs,verbatim, lipsum}
\usepackage{bm}
\usepackage{color}

\begin{document}

\title{Cryogenic Microwave Filter Cavity with a Tunability Greater than 5 GHz}

\author{T.J. Clark}\email{tjclark@ualberta.ca}
\author{V. Vadakkumbatt}
\author{F. Souris}
\author{H. Ramp}
\author{J.P. Davis}\email{jdavis@ualberta.ca}
\address{Department of Physics, University of Alberta, Edmonton, Alberta, Canada T6G 2E9}
%\date{\today}

\begin{abstract}
A wide variety of applications of microwave cavities, such as measurement and control of superconducting qubits, magnonic resonators, and phase noise filters, would be well served by having a highly tunable microwave resonance.  Often this tunability is desired \textit{in situ} at low temperatures, where one can take advantage of superconducting cavities.  To date, such cryogenic tuning while maintaining a high quality factor has been limited to $\sim500$ MHz.  Here we demonstrate a three-dimensional superconducting microwave cavity that shares one wall with a pressurized volume of helium. Upon pressurization of the helium chamber the microwave cavity is deformed, which results in \textit{in situ} tuning of its resonant frequency by more than 5 GHz, greater than 60\% of the original 8 GHz resonant frequency. The quality factor of the cavity remains approximately constant at $\approx7\times 10^{3}$ over the entire range of tuning. As a demonstration of its usefulness, we implement a tunable cryogenic phase noise filter, which reduces the phase noise of our source by approximately 10 dB above 400 kHz.
 
\end{abstract}

\maketitle

Three dimensional (3D) microwave cavities have proven to be an indispensable part of many quantum systems. They have recently been used in conjunction with superconducting qubits\cite{Paik11} to make spectacular progress in the field of cavity QED, where they are integral components in the implementation of universal gate sets,\cite{Heeres17} nondestructive measurement of single microwave photons,\cite{Kono18} and implementation of programmable interference between quantum memories.\cite{Gao18} When paired with ferromagnetic spheroids,\cite{Bai17} they readily exhibit strong coupling to spin resonances,\cite{Zhang14,Maier17} which has allowed for the field of cavity magnonics to flourish with applications such as bidirectional microwave-optical conversion \cite{Hisatomi16} and resolving magnon number states.\cite{Lachance17} When coupled to a mechanical element, such as a membrane \cite{Yuan15} or superfluid helium, \cite{DeLorenzo17} they form an optomechanical system with the potential for exceedingly high cooperativities and quality factors, allowing for the rich toolbox of optomechanics to be employed.

In all of these applications, cryogenic tunability of the microwave cavity -- without sacrificing the quality factor ($Q$) -- would allow for the use of superconducting materials, desirable for the high quality factors they confer. For instance, one can imagine a superconducting qubit encapsulated inside a superconducting resonator. Normally, such a qubit is tuned by the application of a magnetic field, \cite{Devoret04} but this is prohibited inside a superconducting cavity due to field exclusion from the Meissner effect. With a highly-tunable cavity, one could enjoy all the advantages of a superconducting cavity while still being able to control its detuning with respect to a qubit.  Similarly for cavity magnonics, a highly-tunable cavity could allow control of the phonon-magnon hybridization without the need for a tunable magnetic field.

In addition to their uses outlined above, microwave cavities are useful in both classical and quantum systems as phase noise filters.\cite{Pozar09,Rocheleau09,Teufel11} Classically, phase noise filters can be used to improve the data transmission rate in protocols like phase shift keying, where phase noise imparts bit flip errors.\cite{Proakis07} In systems such as the optomechanical one in Ref.~\citenum{DeLorenzo17}, a filter could reduce the base phonon occupancy when sideband cooling, as phase noise causes unwanted cavity population, which in turn increases the phonon occupancy.

Prior implementations of tunable cavities have employed mechanical,\cite{LeFloch13} piezoelectric,\cite{Carvalho14,Carvalho16} or capacitive \cite{Liu10} actuation to shift either a wall of the cavity or a re-entrant stub, or have introduced dielectric or metallic materials into the cavity volume to change the effective permittivity of the system.\cite{Stammeier17} Both of these approaches have been successful, and are able to bring about useful frequency shifts, but have inherent issues when applied at cryogenic temperatures. Piezoelectric crystals become much less active at cryogenic temperatures, which cripples the tunability of a cavity tuned by them. \cite{Carvalho14} The insertion of external materials necessitates complicated actuation stages at cryogenic temperatures that are difficult to implement.  Here we show that a highly-tunable, cryogenically-compatible, superconducting microwave cavity is possible.  We demonstrate a tunability greater than 5 GHz -- the largest cryogenic tunability demonstrated to date -- with a quality factor of $\approx7\times10^3$. As an example, we employ our cavity as a phase noise filter and measure a 10 dB reduction in phase noise above 400 kHz.  

Our design uses an aluminum  membrane (0.9 mm) brought within close proximity to a re-entrant stub, similar to the designs of Refs.~\citenum{Carvalho14,Carvalho16}. The two halves of the aluminum microwave cavity were machined flat and polished.  The stub was machined to be 70 $\mu$m away from the flat membrane. As can be seen in the finite element simulation in Fig.~1, this design localizes the electric field to the regime between the stub and the membrane. The aluminum membrane is backed by a volume of liquid helium, which is fed by a thermally-anchored capillary.\cite{Souris17} This volume, when pressurized, deforms the membrane similar to the classic Straty-Adams pressure gauge,\cite{Straty69} and results in the aforementioned large tunability.  One advantage of this design over our previous work \cite{Souris17} is that -- in addition to the larger tunability -- no specialty hermetic SMA feedthroughs are needed.  Instead, the low-temperature-leak-tight liquid helium chamber and microwave cavity share only the aluminum membrane, greatly reducing the fabrication complexity.  Two loop couplers \cite{Reagor16} are inserted into the cavity, which enables transmission measurements, as well as the use of the cavity as a filter. The depth of these coupling loops is chosen such that both loops are approximately critically coupled to the cavity, which maximized the power transferred through the cavity.

All of our measurements are carried out with our cavity thermally anchored to the base-plate of a dilution refrigerator, as shown in Fig.~1, and cooled to $\sim200$ mK. The cavity is stimulated through $\approx40$ dB of thermalizing attenuation, and the cavity throughput is amplified by 37 dB via a HEMT amplifier attached to the 4 K stage, before room-temperature measurement. In order to characterize the tunability of our cavity, we interrogate the system with a vector network analyzer (VNA) and record the transmission scattering parameter (S21) at a variety of helium pressures. The measurement schematic is outlined in Fig.~1.

\begin{figure}[t]
\includegraphics[width= 0.45\textwidth]{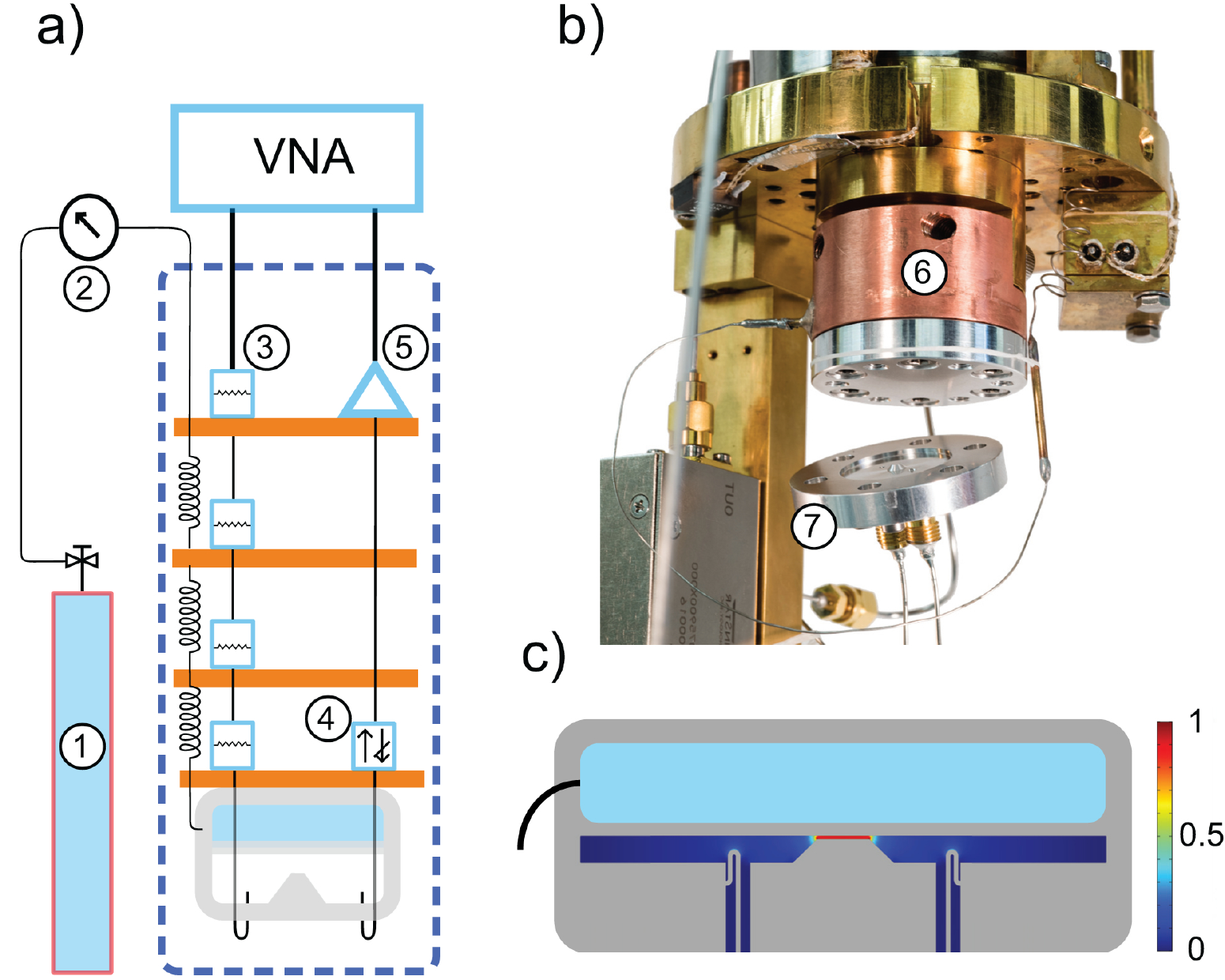}
\caption{a) A schematic outlining the system, showing the (1) helium reservoir and (2) pressure control, as well as the (3) attenuated microwave coax, (4) isolator, and (5) HEMT amplifier, inside of the cryogenic environment.  The tunable cavity is mounted to the base-plate of the dilution refrigerator.  b) A photograph of the base of our dilution refrigerator. The helium volume is contained in the copper cell (6).  The polished aluminum cavity (7) is shown opened, to expose the bottom of the membrane and the re-entrant stub. c) A finite element simulation of the electric field within the cavity for the resonance used here, normalized to the electric field between the movable membrane and the stub.  The capacitance at this position is highly sensitive to the displacement of the membrane, resulting in large shifts in the cavity resonance frequency with small changes in the membrane position.}
\end{figure}

\begin{figure}[b]
\includegraphics[width=0.45\textwidth]{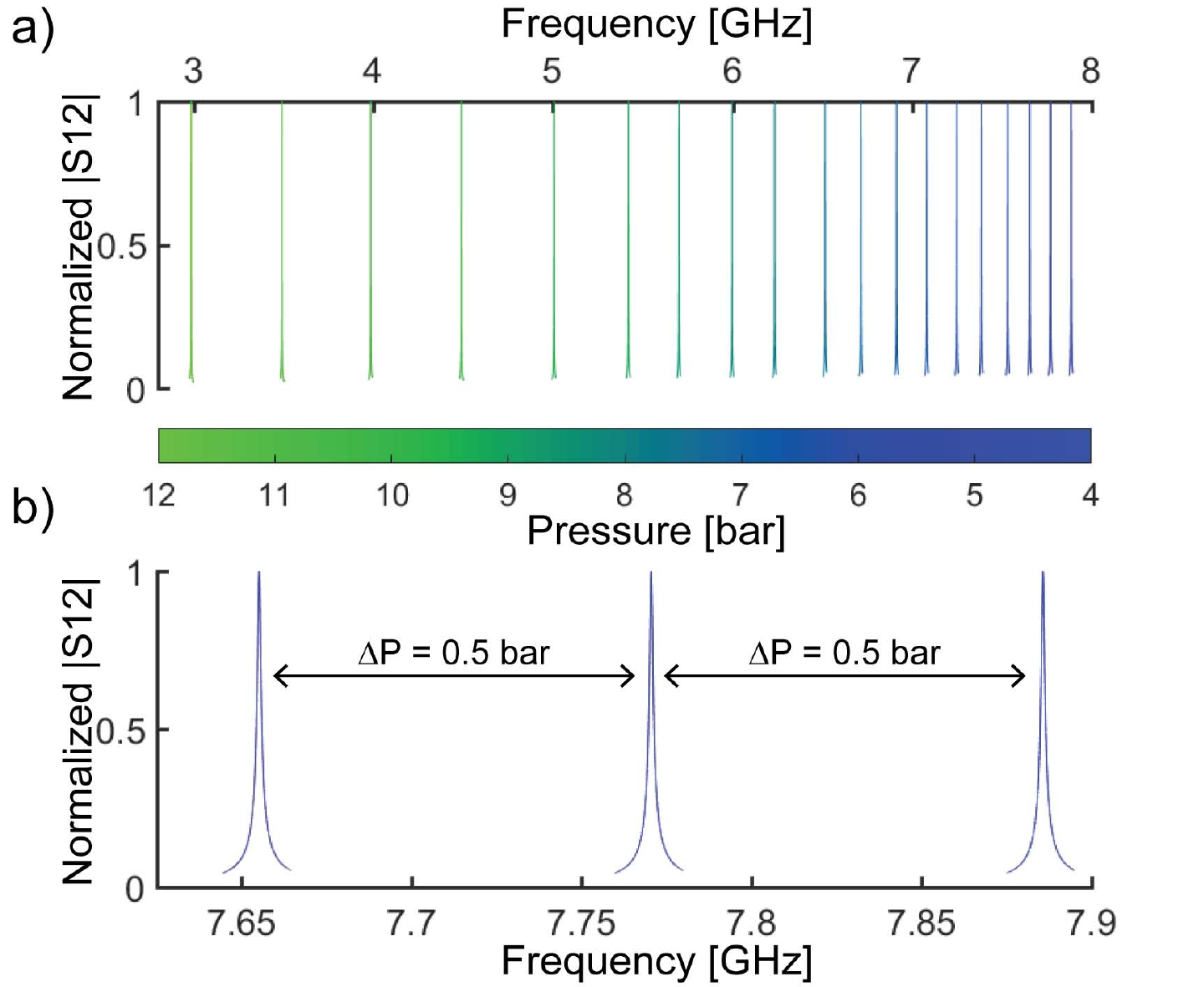}
\caption{
a) Normalized cavity resonances, each separated by 0.5  bar, displaying the large tunability of our cavity at cryogenic temperatures. Greater than 5 GHz of tunability is evident, limited only by our measurement electronics. }
\end{figure}

The large tunability of this cavity can be understood intuitively by considering the capacitive contribution of the membrane to the resonance frequency of the 3D cavity. The membrane can be roughly approximated as a parallel plate capacitor, with capacitance $C = \epsilon_0 A/d$ where $\epsilon_0,A,d$ are the permittivity of free space, the capacitor area, and the distance between the plates of the capacitor respectively. As the membrane deforms, the capacitance changes, as does the microwave resonant frequency. This frequency is given by $\omega = {1}/\sqrt{LC}$ where $L$ is the inductance of the resonator and $C$ is its capacitance. 

More accurately, the displacement of the membrane can be described by the deformation relation for a circularly clamped plate under uniform loading.\cite{Young17} This relation predicts that the average displacement of the membrane responds linearly with applied pressure. Additionally, the cavity resonant frequency can be described by the general expressions for re-entrant cavities. \cite{Fujisawa58,LeFloch13} 
These expressions can be combined to produce an expression for the resonant frequency of our cavity as follows:

\begin{equation}
     \omega_c = \frac{1}{2\pi} \sqrt
     \frac{\mu_0\epsilon_0}
     {r_\textrm{post}h \bigg( \frac{r_\textrm{post}}{2(d-XP)} +
     \frac{2}{\pi}\log\left(\frac{el_m}{(d-XP)}\right)  \bigg)
     \log\left(\frac{r_\textrm{cav}}{r_\textrm{post}} \right)},
\end{equation}
\normalsize
where $r_\textrm{post}$, $r_\textrm{cav}$, and $h$ are the radius of the post, the radius of the cavity, and the height of the cavity respectively. $X$ is a responsivity parameter describing the sensitivity of the membrane to applied pressure, $P$, that depends on geometry and material properties.\cite{Young17} $l_m$ is a purely geometric factor, given for our geometry by $l_m = 0.5 \sqrt{(r_\textrm{cav}-r_\textrm{post})^2 +h^2}$, and $e$ is Euler's constant. In either scenario, one can see that as the pressure increases and the gap between stub and membrane becomes small, the capacitance stemming from the gap between the stub and the membrane changes rapidly, which translates into large changes in the resonant frequency.

In Fig.~2 we show a series of normalized cavity resonances as the pressure is varied isothermally. The nonlinear frequency tuning, growing more responsive at higher pressures, is consistent with the above expectations. Fig.~2 also presents a closer view of the cavity responses at low pressure, demonstrating their Lorentzian shape more clearly. In order to determine the cavity parameters $Q$ and $f_0$ to characterize the tunability of our cavity and its expected filter cutoff frequency, we fit our complex scattering data to the expected shape of a cavity resonator in transmission. \cite{Probst15,Reagor16}

Fig.~3 displays these extracted parameters, as well as a fit of Eq.~1 to the resonant frequency as a function of pressure. This fit allows us to extract an experimental value for both the responsivity parameter $X$, and the gap between the re-entrant stub and our membrane $d$. We find that the membrane has a responsivity of 7.3 $\mu$m/bar, and that the gap between the membrane and the stub is 100 $\mu$m, which is close to the design goal of 70 $\mu$m. It is worthy of note that the fit is imperfect. This imperfection likely comes from a deviation of the assumption of small membrane deflection in the deformation relation for circularly clamped plates, and from the inaccuracy imparted from the small ratio of $r_\textrm{cav} /l_m$ as described in Ref.~\citenum{Fujisawa58}. More importantly, while the cavity frequency changes significantly, the $Q$ is largely unaffected, decreasing slightly at only the highest pressures. The modest $Q$s achieved in the present architecture likely result from the seam losses, that is from the current flowing across the junction between the cap and body of the cavity.\cite{Cohen17} These losses could be reduced by the use of high purity aluminum and careful surface treatment,\cite{Reagor13} or by electroplating with indium.\cite{Brecht15}

\begin{figure}[b]
\includegraphics[width=0.4\textwidth]{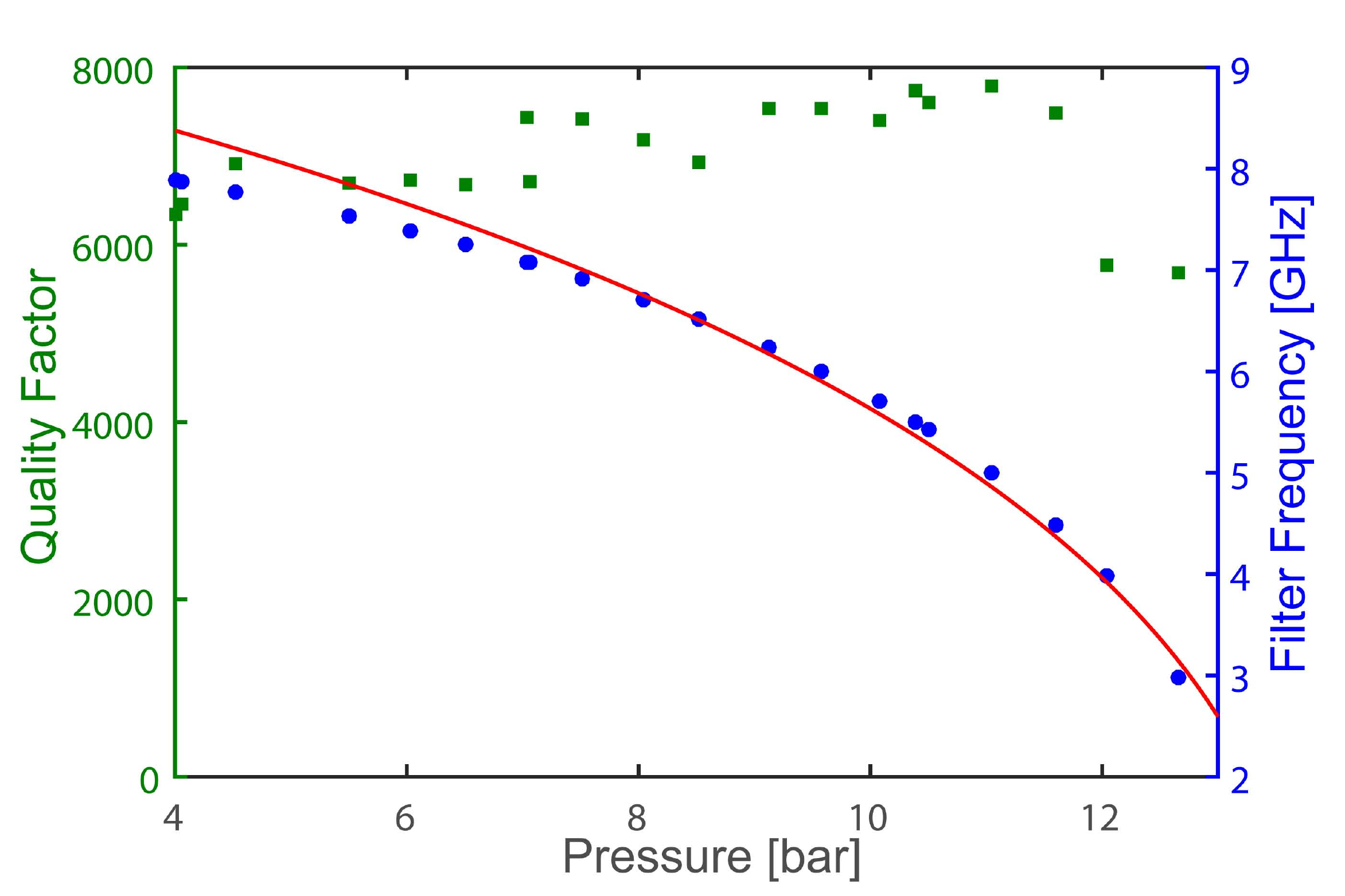}
\caption{Quality factors and center frequencies extracted from the data in Fig.~3. A fit to frequencies using Eq.~1 is overlaid in red, from which the responsivity parameter ($X = 7.3~\mu$m/bar) and stub-membrane distance ($d = 100~\mu$m) are extracted. The quality factor remains approximately constant except for at the highest pressures, and the resonant frequency follows the curve, with deviations explained in the main text. }
\end{figure}

With the tunability of the cavity understood and demonstrated, we now turn our attention to the cavity's utility as a tunable phase noise filter. We have constructed a delay line discriminator (DLD) to characterize the cavity's effectiveness as a phase noise filter.\cite{Packard85} Our apparatus for this measurement is outlined in Fig.~4. To understand the operation of a DLD, one begins with consideration of the phase noise as a frequency jitter, an equivalence that holds as long as the perturbations to phase are small. A noisy signal to be measured is split into two waveguides, one much longer than the other, but with their length tuned such that the phase difference between the two arms at the test frequency is zero. As noise causes deviation from this frequency, the relative phase in the arms is shifted from zero due to the different lengths of waveguide. This phase shift is transduced into a voltage by a mixer, and this voltage is subsequently Fourier transformed. The magnitude of this voltage spectral density is a double-sided measure of the phase noise. It is important to note that the sensitivity of the DLD scheme is poor at low frequencies, which explains the quantitative disagreement with the manufacturer's specifications below  $\sim1$ MHz.\cite{Packard85}

\begin{figure}[t]
\includegraphics[width=0.48\textwidth]{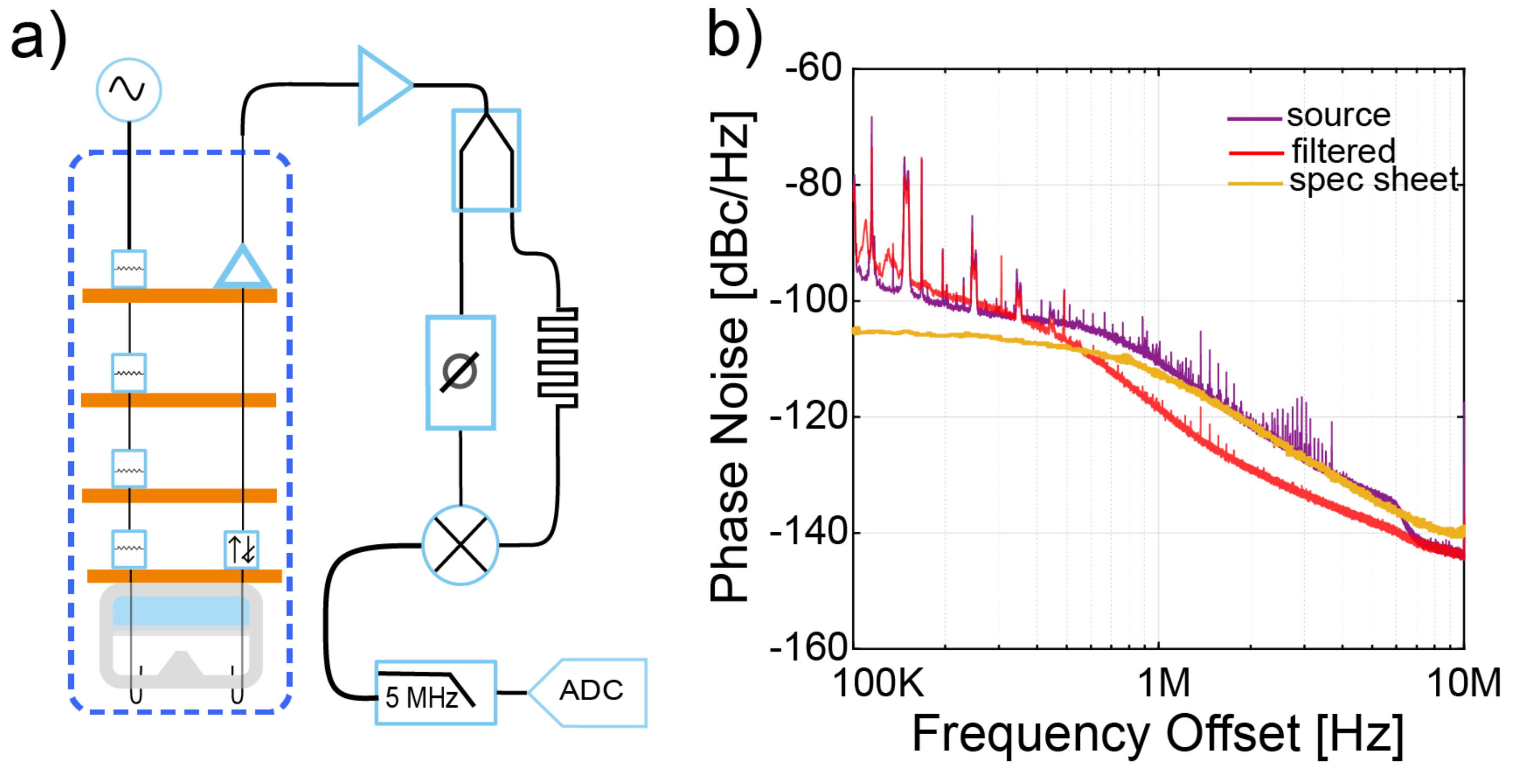}
\caption{a) Schematic outlining the measurement apparatus used to demonstrate our cavity's phase noise filtration. In contrast to Fig.~1, the measurement is executed using a home-built delay line discriminator (described in main text) and recorded using an analog-to-digital convertor (ADC).  b) Phase noise of our signal generator at 6 GHz, superimposed on the phase noise of the same generator in line with our filter cavity. Clear filtration is evident above 400 kHz, consistent with expectations from the cavity $Q$. The drop in noise above 5 MHz is a result of the low-pass filter used before the ADC.  Data from the specification sheet of the source (yellow trace) at 6 GHz confirms the accuracy of our DLD phase noise measurement.}
\end{figure}

We measure the phase noise of both our bare source (SRS SG384), and our source after filtration, with our tunable filter. Comparison to the specifications of the source verifies the accuracy of our measurement scheme, as shown in Fig.~4b.  The drop-off in the source phase noise at 6 MHz is a result of the 5 MHz low-pass filter used before the ADC to remove the high frequency components of the mixer output, as shown in Fig.~4a. The absence of this drop-off in the filtered data indicates that the filtered measurement is already at the noise floor of our measurement scheme. This, in conjunction with the expectations for phase noise filtration outlined in Ref.~\citenum{DeLorenzo16}, suggests that our filter rejects even more noise than is indicated in this data, but this additional filtration is obscured by our measurement noise-floor.  It is also noteworthy the presence of a forest of spikes in the phase noise, not shown in the manufacturers specifications.  These are real, reproducible, phase noise peaks that are visible because of the high resolution bandwidth of our measurement. These phase noise peaks are dramatically reduced via our cryogenic cavity filter, as seen in Fig.~4b, a testament to the utility of such a tunable microwave cavity filter.

In conclusion, we have demonstrated a cryogenically-compatible 3D microwave cavity that exhibits more than 5 GHz of \textit{in situ} tunability without serious reduction in the quality factor.  We have explicitly demonstrated its use as a phase noise filter, where we have shown its ability to remove $\sim10$ dB of phase noise from a commercial microwave source. Furthermore, we have described a simple, inexpensive method to measure this filtration. 

The unprecedented tunability of our cavity at cryogenic temperatures has potential applications as both an enabling technology for, and as an integral component of, various advanced microwave systems. As demonstrated here, our cavity can be used to reduce the phase noise of a microwave source, which has applications for a variety of electromechanical and cavity QED experiments. In addition to this, one can imagine using our cavity as the microwave resonator in a cavity magnonic or electromechanic system, where its unique tunability would enable magnetic-field-free manipulation of the detuning between the cavity and a coupled resonator.  Finally, a system like this one but constructed from a non-superconducting material could potentially be used as a tunable cavity attenuator.\cite{Wang18} Such an attenuator has been shown to thermalize the drive line of a superconducting qubit better than traditional resistive attenuators, leading to improved coherence times.\cite{Wang18} Employment of our design could provide not only an improvement in the coherence time due to the increased thermalization, but also impart a time-varying coupling between the qubit and its environment that could be used to even further enhance this coherence. 

This work was supported by the University of Alberta, Faculty of Science; the Natural Sciences and Engineering Research Council, Canada (Grants Nos. RGPIN-2016-04523, DAS492947-2016, and STPGP 494024 - 16); and the Canada Foundation for Innovation.

\end{document}